\documentclass[journal]{IEEEtran}  % Comment this line out if you need a4paper

\IEEEoverridecommandlockouts                              % This command is only needed if 
\usepackage{graphics} % for pdf, bitmapped graphics files
\usepackage{epsfig} % for postscript graphics files
\usepackage{amsmath} % assumes amsmath package installed
\usepackage{amssymb}  % assumes amsmath package installed
\usepackage{amsthm}  % assumes amsmath package installed
\usepackage{enumerate}

\usepackage{afterpage}
\usepackage{euscript}
\usepackage{multirow}
\usepackage{xparse}% http://ctan.org/pkg/xparse
\usepackage{cite}

\NewDocumentCommand{\overarrow}{O{i^{th}} O{\uparrow} m}{%
  \overset{\makebox[0pt]{\begin{tabular}{@{}c@{}}#3\\[0pt]\ensuremath{#2}\end{tabular}}}{#1}
}

\usepackage{mathtools}
\title{\LARGE \bf
Knee Compliance Reduces Peak Swing Phase Collision Forces in a Lower-Limb Exoskeleton Leg: A Test Bench Evaluation
}

\author{Stefan O. Schrade, Marcel Menner, Camila Shirota, Peter Winiger, Alex Stutz, Melanie N. Zeilinger, Olivier Lambercy and Roger Gassert% <-this % stops a space
        \thanks{S.O. Schrade, C. Shirota, O. Lambercy, and R. Gassert are with the Rehabilitation Engineering Laboratory,
        ETH Zurich, 8092 Zurich, Switzerland   \hspace{3cm}
        {\tt\small \{stefan.schrade,camila.shirota,\newline olivier.lambercy,roger.gassert\}@hest.ethz.ch}}
        \thanks{M. Menner and M.N. Zeilinger are with the Institute for Dynamic Systems and Control,
        ETH Zurich, 8092 Zurich, Switzerland   \hspace{3cm}
        {\tt\small \{mmenner,mzeilinger\}@ethz.ch}}
        \thanks{A. Stutz and P. Winiger are with the Laboratory for Mechanical Systems Engineering, Swiss Federal Laboratories for Materials Science and Technology
        EMPA D\"ubendorf, 8600 D\"ubendorf, Switzerland   \hspace{3cm}
        {\tt\small \{peter.winiger,alex.stutz\}@empa.ch}}
}

\usepackage{pstool}

\begin{document}

\newcommand{\DataPath}{"data"}
\newcommand{\PlotPath}{""}

\maketitle
\thispagestyle{empty}
\pagestyle{empty}

%%%%%%%%%%%%%%%%%%%%%%%%%%%%%%%%%%%%%%%%%%%%%%%%%%%%%%%%%%%%%%%%%%%%%%%%%%%%%%%%
% ------ ABSTRACT --------------------------------------------------------------
%%%%%%%%%%%%%%%%%%%%%%%%%%%%%%%%%%%%%%%%%%%%%%%%%%%%%%%%%%%%%%%%%%%%%%%%%%%%%%%%

\begin{abstract}
Powered lower limb exoskeletons are a viable solution for people with a spinal cord injury to regain mobility for their daily activities. However, the commonly  employed rigid actuation and pre-programmed trajectories increase the risk of falling in case of collisions with external objects. Compliant actuation may reduce forces during collisions, thus protecting hardware and user. However, experimental data of collisions specific to lower limb exoskeletons are not available. In this work, we investigated how a variable stiffness actuator at the knee joint influences collision forces transmitted to the user via the exoskeleton. In a test bench experiment, we compared three configurations of an exoskeleton leg with a variable stiffness knee actuator in (i)~compliant or (ii)~stiff configurations, and with (iii)~a rigid actuator. The peak torque observed at the pelvis was reduced from 260.2~Nm to 116.2~Nm as stiffness decreased. In addition, the mechanical impulse was reduced by a factor of three. These results indicate that compliance in the knee joint of an exoskeleton can be favorable in case of collision and should be considered when designing powered lower limb exoskeletons. Overall, this could decrease the effort necessary to maintain balance after a collision and improved collision handling in exoskeletons could result in safer use and benefit their usefulness in daily life.
\end{abstract}

%%%%%%%%%%%%%%%%%%%%%%%%%%%%%%%%%%%%%%%%%%%%%%%%%%%%%%%%%%%%%%%%%%%%%%%%%%%%%%%%
% ------ KEYWORDS --------------------------------------------------------------
%%%%%%%%%%%%%%%%%%%%%%%%%%%%%%%%%%%%%%%%%%%%%%%%%%%%%%%%%%%%%%%%%%%%%%%%%%%%%%%%


%%%%%%%%%%%%%%%%%%%%%%%%%%%%%%%%%%%%%%%%%%%%%%%%%%%%%%%%%%%%%%%%%%%%%%%%%%%%%%%%
% ------ INTRODUCTION ----------------------------------------------------------
%%%%%%%%%%%%%%%%%%%%%%%%%%%%%%%%%%%%%%%%%%%%%%%%%%%%%%%%%%%%%%%%%%%%%%%%%%%%%%%%

\section{Introduction}
Many people are limited in their mobility after surviving a spinal cord injury (SCI). Wheelchairs are the state-of-the-art solution to regain mobility if ambulation is difficult or no longer possible. Aside from challenged mobility, the SCI population suffers from increased risk of experiencing secondary health problems such as cardio-pulmonary diseases, obesity and musculoskeletal degeneration~\cite{Walter2002}. Some of these are thought to be fostered by the inability to stand and walk~\cite{Eng2001,Arva2009}. Hence, powered lower limb exoskeletons could mitigate some of these effects as they allow the user to regain the ability to walk. Such devices are available for use in clinics (e.g.~Ekso and EksoGT, Ekso Bionics, USA)~\cite{Kolakowsky2013,BachBaunsgaard2018} and also home environments (ReWalk Personal~6.0 and ReWalk Rehabilitation, ReWalk~Robotics, USA)~\cite{Esquenazi2012,Zeilig2012}.

Powered lower limb exoskeletons are attached to the legs, pelvis and sometimes torso of the user. The exoskeleton moves the user's biological legs with its mechanical legs while supporting the upper body. Most devices are anthropomorphic and actuate hip and knee joints in the sagittal plane, flexing and extending these joints. Crutches or other walking aids are used for balancing and to compensate for missing active ankle joints. Typically, exoskeletons are actuated with electric motors and high reduction ratio gearboxes, which enable high torques and precise position control. The movements are pre-programmed fixed position or velocity controlled trajectories~\cite{Schrade2018,Griffin2017,Vouga2017} or, more seldomly, torque controlled trajectories~\cite{Oh2015}. Since most movement patterns were designed for even terrain, rigid actuation in combination with the fixed patterns can lead to unsuitable control in unstructured environments as those encountered during daily life activities. The exoskeleton can typically not detect obstacles and may collide with them during swing, or force the leg to further extend on uneven ground despite already having established ground contact. In both cases, the user will likely be pushed backwards, leading to uncomfortable or even unsafe situations. In the worst case, falls may be induced if the user cannot compensate the inappropriate control with the walking aids. 

A particularly challenging and potentially dangerous situation is when an obstacle hinders forward leg swing during exoskeleton walking. Noticing and reacting to such collisions is especially challenging for users with motor complete SCI as they often also have impaired sensory perception. In contrast, unimpaired humans are capable of quickly and smoothly stepping over obstacles or prematurely terminating the swing phase to regain balance~\cite{Schillings2000}. Current exoskeletons, however, are not able to detect and react to collisions.

The robotics community has studied collisions~\cite{Zheng1985}, added collision avoidance~\cite{Brunn1996} as well as detection~\cite{DeLuca2006,Haddadin2008}, and implemented mitigation strategies~\cite{DeLuca2006,Haddadin2008} to address this issue. Dealing with collisions becomes ever more important as robots leave well-controlled and structured factory environments, as is the case for exoskeletons.

Rendering joints compliant has been suggested to reduce the severity of impacts~\cite{Haddadin2007,Haddadin2007a,Park2008}. This can be achieved either with mechanical designs or through control. If compliance is emulated through control, sensing and actuation could be too slow or not robust enough to react to the fast dynamics of collisions~\cite{Haddadin2007}, and high peak forces that can damage the hardware and may be dangerous to users. Hence, implementing a mechanical compliance may be more desirable. This has been suggested for assistive devices for walking~\cite{Cestari2014, Cherelle2010, Bacek2017}.

However, the influence of compliance in joints has not yet been experimentally investigated. In general, robot collision experimental data are scarce. Povse et al.~\cite{Povse2010} investigated damage to human soft tissue that may be induced from collisions between humans and robots. Haddadin~et~al.~\cite{Haddadin2007} took the approach one step further and gave the robot sharp tools to investigate how collisions between robots and humans can be rendered safe. Park et al.~\cite{Park2008} found that a nonlinear spring integrated in the robot joint may significantly reduce the risk of head injury in case of a collision between a robot arm and a human head. However, none of these studies are specific to the application of lower limb exoskeletons. 

Simulation results~\cite{Haddadin2009} motivated our hypothesis that the peak forces experienced by the user would be decreased in amplitude and delayed with a soft joint in comparison to a rigid one. Decreased force amplitude could reduce the need to compensate for collision forces and torques, which could simplify keeping balance. Delayed peak forces would give more time to the user and controller to react to collisions.

Thus, the goal of this work was to investigate the influence of joint compliance on the resulting interaction forces and torques for lower limb exoskeletons. We investigated swing phase collisions with one leg of the VariLeg exoskeleton~\cite{Schrade2018} on a test bench setup using three different stiffness levels: two were rendered with the VSA and the third, representing a conventional rigid joint, by blocking the VSA. As outcome parameters, we measured the collision force at the obstacle, the torque and the resulting force acting at the attachment point of the leg to the test frame at the pelvic level in the plane of movement (sagittal plane). The latter were motivated by the assumption that the forces and torques at the pelvic level would be comparable to the forces and torques that would be transmitted to a user of the exoskeleton.

%%%%%%%%%%%%%%%%%%%%%%%%%%%%%%%%%%%%%%%%%%%%%%%%%%%%%%%%%%%%%%%%%%%%%%%%%%%%%%%%
% ------ METHODS ---------------------------------------------------------------
%%%%%%%%%%%%%%%%%%%%%%%%%%%%%%%%%%%%%%%%%%%%%%%%%%%%%%%%%%%%%%%%%%%%%%%%%%%%%%%%

\section{Methods}
\subsection{The VariLeg Exoskeleton Leg}
\begin{figure*}[h!tb]
    \centering
    \includegraphics[width=\textwidth]{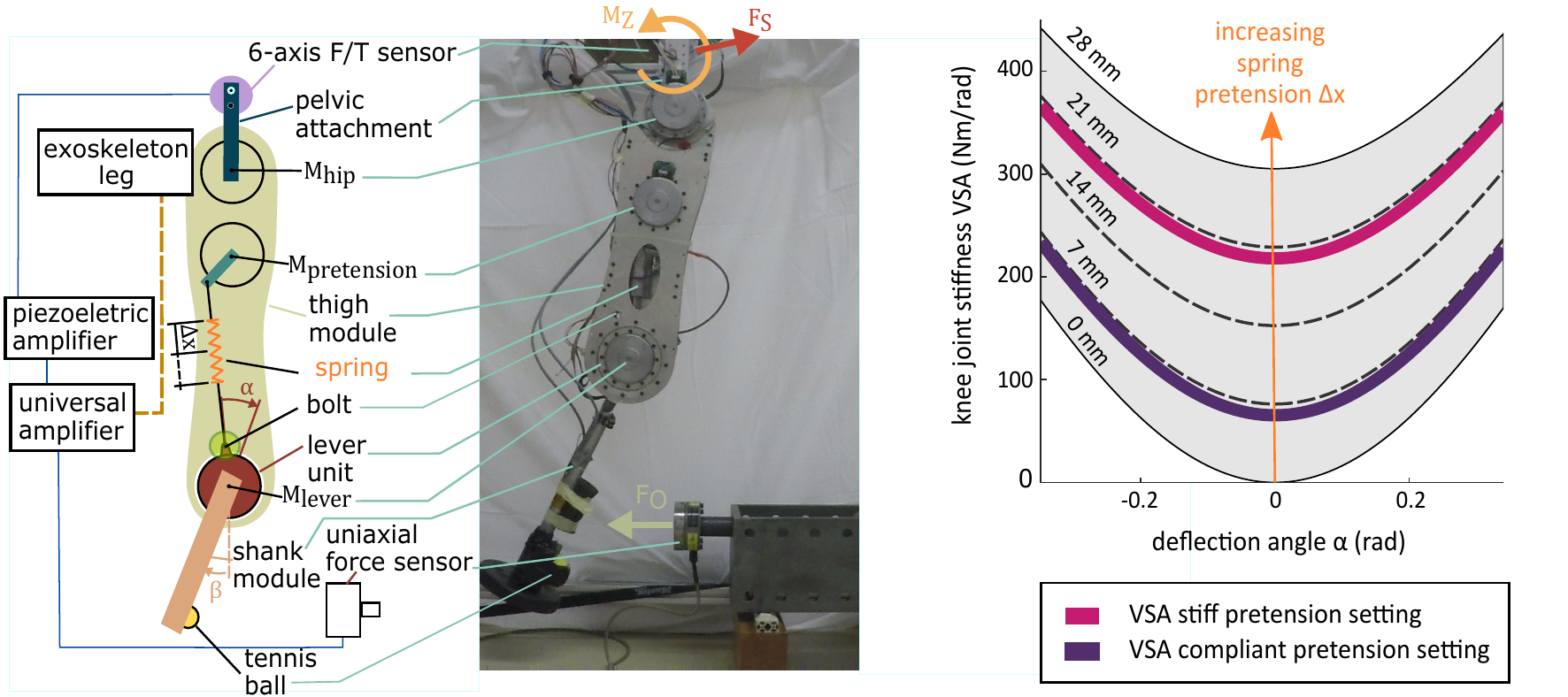}
    \caption[Overview of the setup for the collision experiments]
    {\textbf{Overview of the setup for the collision experiments} The Variable Stiffness Actuator (VSA) can be blocked by inserting a bolt (left; green circle). If the VSA is active (not blocked), the pretension motor ($M_{pretension}$) controls joint stiffness by varying the pretension of the spring (right). Data were recorded from the universal amplifier and the exoskeleton. A shared trigger signal was used to synchronize the data logs of the exoskeleton and the universal amplifier. Two VSA configurations were used: VSA STIFF and VSA COMPLIANT (right). The collision force $F_O$ was measured at the obstacle. At the pelvic level the sagittal plane force $F_S$ and the pitch torque $M_Z$ were measured during the collision.}
    \label{fig:exerimental_setup}
\end{figure*}

The VariLeg is a powered lower limb exoskeleton designed to restore the walking ability of motor-complete spinal cord injured users~\cite{Schrade2018}. The unique feature of the exoskeleton is the Variable Stiffness Actuator (VSA) in each knee joint~(Fig.~\ref{fig:exerimental_setup}). These actuators cover a stiffness range similar to the human knee joint~\cite{Pfeifer2014}, and allow varying it online during walking. The VSA can also be blocked with a bolt, resulting in a conventionally actuated rigid knee joint. The exoskeleton also has conventional rigid actuators that allow hip flexion and extension~($M_{hip}$). For the test bench experiments, the passive spring-loaded ankle was replaced with a dumb-bell disk of equivalent to the weight of the exoskeleton ankle and foot~(1.5~kg) and a tennis ball substituting the semi-compliant shoe tip and foot. The exoskeleton has the capability of varying the knee joint stiffness with a VSA, using the MACCEPA principle~\cite{VanHam2007}. As a consequence, $M_{lever}$ only directly controls the knee angle $\beta$ if a bolt is inserted to connect thigh and shank over the actuator unit. If the bolt is not inserted, the shank and thigh are compliantly connected through the spring. In this case, $M_{lever}$ sets the equilibrium position of the knee angle. Thus, the shank may be deflected from its equilibrium position by angle $\alpha$ if external forces are present. The joint compliance can be varied by pretensioning the spring (i.e. increasing stiffness) with $M_{pretension}$ (Fig.~\ref{fig:exerimental_setup}, right).

For the collision experiments, the leg was suspended from a rigid frame using the attachment point of the pelvic structure of the exoskeleton. A  piezoelectric six-axis force and torque sensor (Type~9306A, Kistler Instrumente AG, Switzerland) was placed between the leg and the frame. The piezoelectric signals were amplified and sampled at 10 kHz by a LabAmp~5167A80 (Kistler Instrumente AG, Switzerland). The obstacle was equipped with a single-axis force sensor (GTM~20kN, GTM Testing and Metrology GmbH, Germany). All data were routed to a universal amplifier (HBM~MX1615, Hottinger Baldwin Messtechnik GmbH, Germany) and sampled at 19.2 kHz. The exoskeleton logged joint angle data at 100 Hz and sent a trigger signal to the universal amplifier indicating when it was recording. This trigger signal was later used to synchronize the data of the universal amplifier with the data of the exoskeleton.

\subsection{Control of the Test Bench Exoskeleton Leg}

\begin{figure}[!h]
      \centering
     \includegraphics[trim={.6cm 0 0 0},clip,width=0.96\columnwidth]{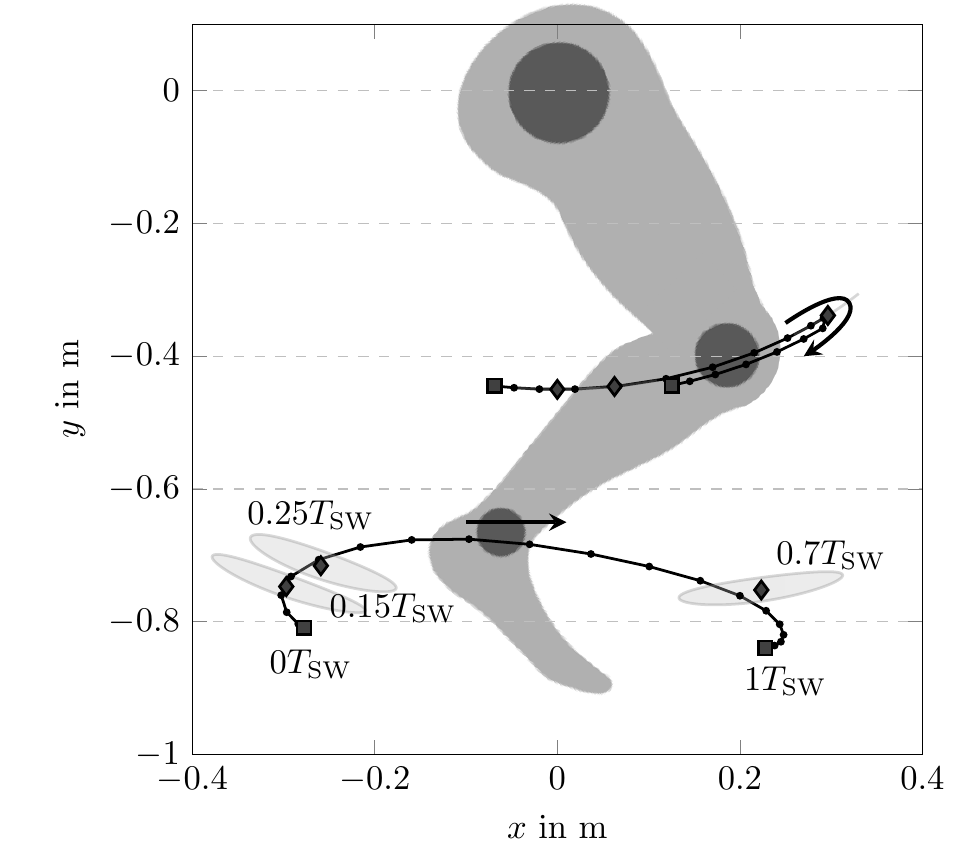}
      \caption[Trajectory generation for the exoskeleton leg]
      {\textbf{Trajectory generation for the exoskeleton leg} The end-point and knee trajectory found by optimization are indicated as curves. Waypoints (diamonds) in the end-point space are specified to be passed within a certain tolerance in position (grey shaded ellipses) and time (specified relative to the desired swing phase time $T_{\rm SW}$ but not visible in this plot). The desired movement starts and ends at the squares. The trajectory of the knee joint is not a design parameter, but rather results from the desired end-point trajectory and the optimization performed to achieve this trajectory.}
      \label{fig:traj_gen}
\end{figure}
To focus on the influence of the differing joint stiffness, we relied on feed-forward control to execute the swing phase. The controller was designed to be the first layer of control inspired by nature, which features systems that inherently behave in a desired way, using mostly feed-forward control with appropriate mechanical structure and without feed-back \cite{Tytell2011} and corresponds to an approach often used in devices with a layered control approach~\cite{Tucker2015}.

The exoskeleton leg was torque controlled for this experiment. A desired trajectory was found by specifying waypoints that the ankle should pass through at specific times of swing phase (Fig.~\ref{fig:traj_gen}). This trajectory was optimized to satisfy the constraints defined by the waypoints, motor speed limits, and motor torque limits. Feed-forward torque commands were then calculated for every joint stiffness configuration based on a model of the exoskeleton leg to execute the desired motion.
This model was derived using Lagrangian mechanics, where the uncertain viscous friction parameters were fitted using experimental data. 
We refer the interested reader to the Appendix for more detailed information (Chapter~\ref{Appendix:Control}).

\subsection{Experimental Protocol}
Collisions were performed during swing phase to simulate stumbling over an obstacle. The leg was brought to the initial position and the execution of the swing phase was triggered via a button. The obstacle was placed such that the collisions occurred at the point of maximum forward velocity during swing phase. This point was identified by analyzing the reference trajectory identified by the optimization~(Fig.~\ref{fig:traj_gen}). After the collision occurred, the system was shut down manually, cutting the motors off from the power supply. This was done to prevent potential damage to the hardware, since the controller had no strategies to mitigate the collision, and since only the time period of the collision was of interest.

Three configurations were tested: Compliant with VSA (COMPLIANT,~65~Nm/rad), stiff with VSA (STIFF,~218~Nm/rad), and VSA blocked (RIGID,~above 20~kNm/rad according to the Harmonic Drive manual\footnote{The stiffness depends on the torque at the joint; for the torque ranges of 0 to 14~Nm, 14~Nm to 48~Nm, and above 48~Nm, the stiffness is approximately 20, 27, 37~kNm/rad, respectively. Source: \textit{Harmonic Drive Reducers Gear Units~SHD-2SH Engineering Data}

Here, we assumed that the exact torque dependent stiffness of the Harmonic Drive was not important, as it was orders of magnitude larger than that of the VSA.}). The STIFF VSA configuration corresponded to a value used in the training sessions described in our previous work~\cite{Schrade2018} where the VSA pretension motor ($M_{pretension}$) was controlled to hold a constant position throughout the whole gait cycle. The COMPLIANT configuration of this experiment also started at this value before triggering the step, and reduced during swing phase to mimic human knee joint stiffness modulation~\cite{Pfeifer2014}. 

The exoskeleton segment lengths were chosen based on typical segment lengths used in training sessions with our users \cite{Schrade2018} (thigh length 0.480~m, shank length 0.445~m). We conducted 15 collisions for each leg configuration.  Video recordings were acquired for visual analysis of the collisions (GoPro HERO5, GoPro Inc., United States of America) at 120 frames per second.

\subsection{Data Analysis}
Data sets that were corrupted due to setting or measurement errors were rejected (n=2). Force and torque data were filtered with a second-order Butterworth low-pass filter at 250~Hz using the $filtfilt$ command of MATLAB~2017b~(MathWorks, United States of America) and signal offset was removed. Collision onset was defined as the point where the force at the obstacle ($F_O$) crossed a threshold of 18~N, based on the sensor noise level. Data from different collisions were aligned in time at collision onset. Individual trajectories were re-sampled at 6.5 kHz and averaged at each time sample. The per sample standard deviation was also calculated to gauge consistency across measurements.

The movement of the end-point in task space was calculated from the joint angles using forward kinematics. The antero-posterior (x-direction) and vertical (y-direction) speeds were obtained by derivation of the sampled position data. The total speed was calculated as:
\begin{equation}
\begin{aligned}
     v_{tot} = \sqrt{{v_x}^2+{v_y}^2}.
\end{aligned}
\end{equation}

The net force in the plane of movement (sagittal plane, $F_S$) and the torque in the sagittal plane (pitch torque, $M_Z$) were evaluated. The net amplitude of the sagittal plane forces was calculated as:
\begin{equation}
\begin{aligned}
     F_S = \sqrt{{F_x}^2+{F_y}^2}.
\end{aligned}
\end{equation}

Peaks in force and torque and their respective time of occurrence were determined within a 150~ms window after the collision as this was assumed to be the relevant time window and similar time windows have been used in other experiments investigating trips~\cite{Shirota2014}. This short time period after collision was not affected by the manual shutdown of the motors. The area under the curve, which corresponds to the mechanical impulse, was also calculated in the same time window for all three knee joint configurations.

The peak force and torque values, their respective time of occurrence and impulse were tested for statistically significant differences. A Kruskal-Wallis test followed by a post hoc multiple comparison test was performed to compare the results across the three different knee joint configurations. The threshold was set to 0.05 for significant differences, stronger significance levels were indicated if below 0.01 and 0.001.

%%%%%%%%%%%%%%%%%%%%%%%%%%%%%%%%%%%%%%%%%%%%%%%%%%%%%%%%%%%%%%%%%%%%%%%%%%%%%%%%
% ------ RESULTS ---------------------------------------------------------------
%%%%%%%%%%%%%%%%%%%%%%%%%%%%%%%%%%%%%%%%%%%%%%%%%%%%%%%%%%%%%%%%%%%%%%%%%%%%%%%%

\section{Results}
\begin{figure}[htb]
    \centering
    \includegraphics[width=\textwidth/2]{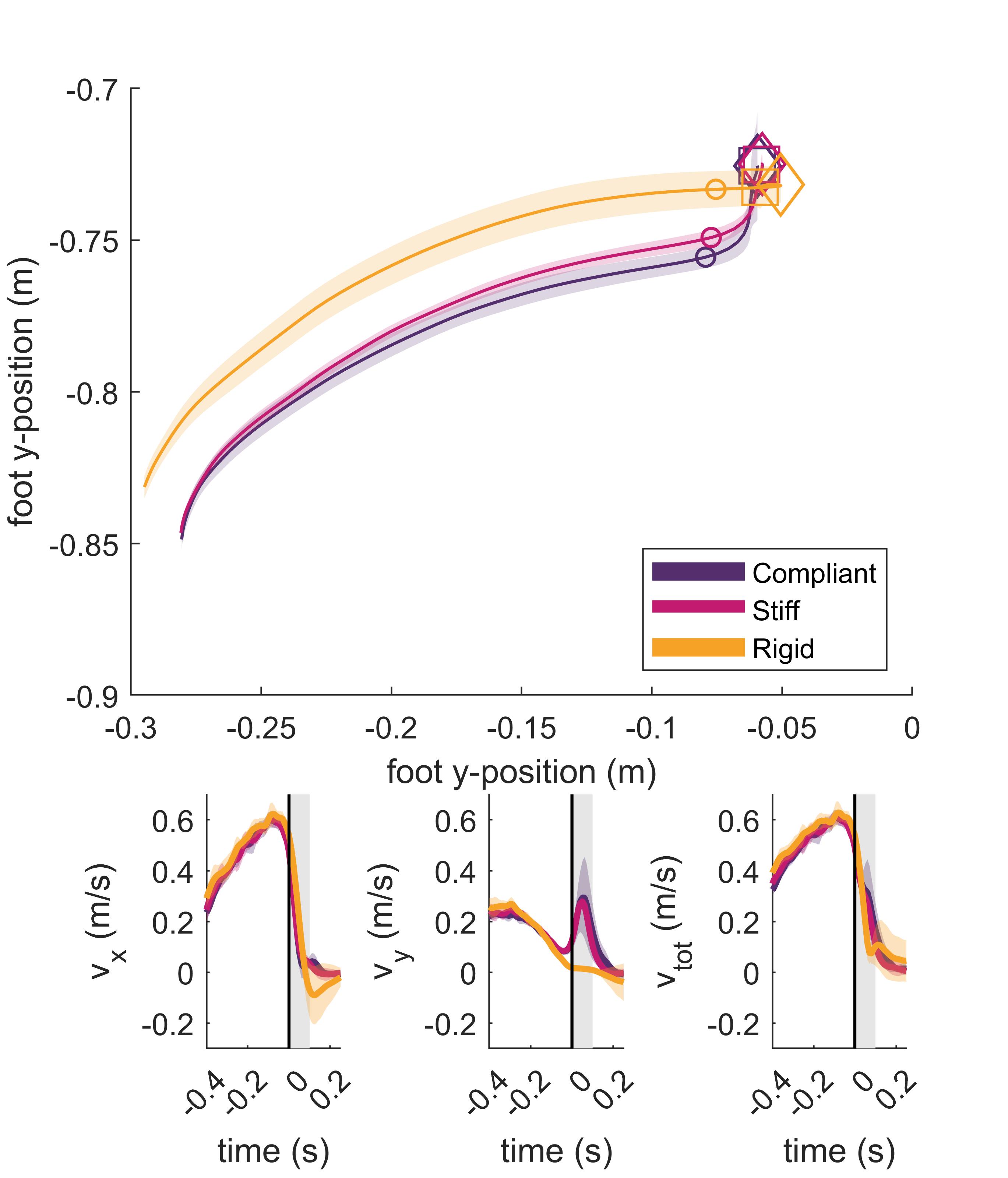}
    \caption[Average end-point position trajectory during swing phase collision]{\textbf{Average end-point trajectory and velocity from beginning of swing phase until the end of the analysis time window} The trajectory for all three configurations started very similar and was consistent (top). Circles indicate location of collision, diamonds indicate maximum position in x and squares indicate position at the end of the analysis window. At the bottom the velocities in different directions are indicated before and after collision. The black line indicates the start of the collision and the grey area marks the analyzed time window of 150~ms. After collision, the RIGID knee joint caused the ankle to bounce off the obstacle, seen as negative x-direction velocity (bottom left) and almost straight average foot trajectory after collision. The COMPLIANT and STIFF joints slid up the surface of the obstacle resulting in an increase in y-direction velocity (bottom middle).}
    \label{fig:exo_signal}
\end{figure}

The trajectory up to the point of collision was similar for all three conditions (Fig.~\ref{fig:exo_signal}~top), with the COMPLIANT and STIFF (VSA) conditions slightly lower than the RIGID condition. Forward velocity at collision~(Fig.~\ref{fig:exo_signal}~bottom left) was 0.52$\pm$0.05~m/s (mean$\pm$standard deviation) for the RIGID knee joint. This was slightly lower than the maximum value observed up to the collision of~0.62~m/s. The COMPLIANT and STIFF knee joints achieved identical forward velocities at collision of 0.44$\pm$0.05~m/s. Upwards velocity (y-velocity) was higher for both VSA collisions than for the RIGID condition. The total velocity, which, when squared, is proportional to the kinetic energy of the leg before collision, was very similar for all three conditions 0.47$\pm$0.04~m/s, 0.47$\pm$0.04~m/s and 0.52$\pm$0.05~m/s (COMPLIANT, STIFF, RIGID).

The collisions with the RIGID knee joint lead to movement only in the x-direction, as the end-point bounced back~(Fig.~\ref{fig:exo_signal}~bottom left and middle). In contrast, when the knee joint was COMPLIANT or STIFF, the hip continued flexing while the foot slid upwards on the obstacle, which can be seen as the increasing y-velocity after the collision for both VSA joint configurations (Fig.~\ref{fig:exo_signal}~bottom middle).

\begin{table}[h]
\caption{
Results of the Statistical Analysis
}
\label{tab:stat_results}
\centering
\begin{tabular}{llllc}
Base Condition & Compared to & p-Value & $\Delta$median & \multicolumn{1}{l}{Unit} \\
\hline
\hline
\multicolumn{5}{l}{\textbf{Peak Obstacle Force, $F_O$}}                             \\
COMPLIANT      & STIFF       & 0.01470 & 72.4           & \multirow{3}{*}{N}       \\
COMPLIANT      & RIGID       & 0.00000 & 93.4           &                          \\
STIFF          & RIGID       & 0.01494 & 21.0           &                          \\
\multicolumn{5}{l}{\textbf{Peak Sagittal Force, $F_S$}}                             \\
COMPLIANT      & STIFF       & 0.18836 & 6.3            & \multirow{3}{*}{N}       \\
COMPLIANT      & RIGID       & 0.00000 & 80.5           &                          \\
STIFF          & RIGID       & 0.00066 & 74.2           &                          \\
\multicolumn{5}{l}{\textbf{Peak Pitch Torque, $M_Z$}}                               \\
COMPLIANT      & STIFF       & 0.01361 & 13.8           & \multirow{3}{*}{Nm}      \\
COMPLIANT      & RIGID       & 0.00000 & 141.1          &                          \\
STIFF          & RIGID       & 0.00550 & 127.2          &                          \\
               &             &         &                & \multicolumn{1}{l}{}     \\
\multicolumn{5}{l}{\textbf{Peak Timing Obstacle Force, $F_O$}}                      \\
COMPLIANT      & STIFF       & 0.34124 & -0.0012        & \multirow{3}{*}{s}       \\
COMPLIANT      & RIGID       & 0.00010 & 0.0456         &                          \\
STIFF          & RIGID       & 0.00000 & 0.0469         &                          \\
\multicolumn{5}{l}{\textbf{Peak Timing Sagittal Force, $F_S$}}                      \\
COMPLIANT      & STIFF       & 0.02873 & -0.0049        & \multirow{3}{*}{s}       \\
COMPLIANT      & RIGID       & 0.00105 & 0.0067         &                          \\
STIFF          & RIGID       & 0.00000 & 0.0116         &                          \\
\multicolumn{5}{l}{\textbf{Peak Timing Pitch Torque, $M_Z$}}                        \\
COMPLIANT      & STIFF       & 0.98080 & 0.0006         & \multirow{3}{*}{s}       \\
COMPLIANT      & RIGID       & 0.00001 & 0.0114         &                          \\
STIFF          & RIGID       & 0.00001 & 0.0108         &                         \\
               &             &         &                & \multicolumn{1}{l}{}     \\
\multicolumn{5}{l}{\textbf{Impulse Obstacle Force, $F_O$}}                          \\
COMPLIANT      & STIFF       & 0.01361 & 5.85           & \multirow{3}{*}{Ns}      \\
COMPLIANT      & RIGID       & 0.00000 & 13.45          &                          \\
STIFF          & RIGID       & 0.00550 & 7.60           &                          \\
\multicolumn{5}{l}{\textbf{Impulse Sagittal Force, $F_S$}}                          \\
COMPLIANT      & STIFF       & 0.01299 & 1.95           & \multirow{3}{*}{Ns}      \\
COMPLIANT      & RIGID       & 0.00000 & 3.10           &                          \\
STIFF          & RIGID       & 0.00603 & 1.14           &                          \\
\multicolumn{5}{l}{\textbf{Impulse Pitch Torque, $M_Z$}}                            \\
COMPLIANT      & STIFF       & 0.01361 & 5.18           & \multirow{3}{*}{Nms}     \\
COMPLIANT      & RIGID       & 0.00000 & 11.24          &                          \\
STIFF          & RIGID       & 0.00550 & 6.06           &                          \\
\end{tabular}
\end{table}

\begin{figure}[htb]
    \centering
    \includegraphics[width=\textwidth/2]{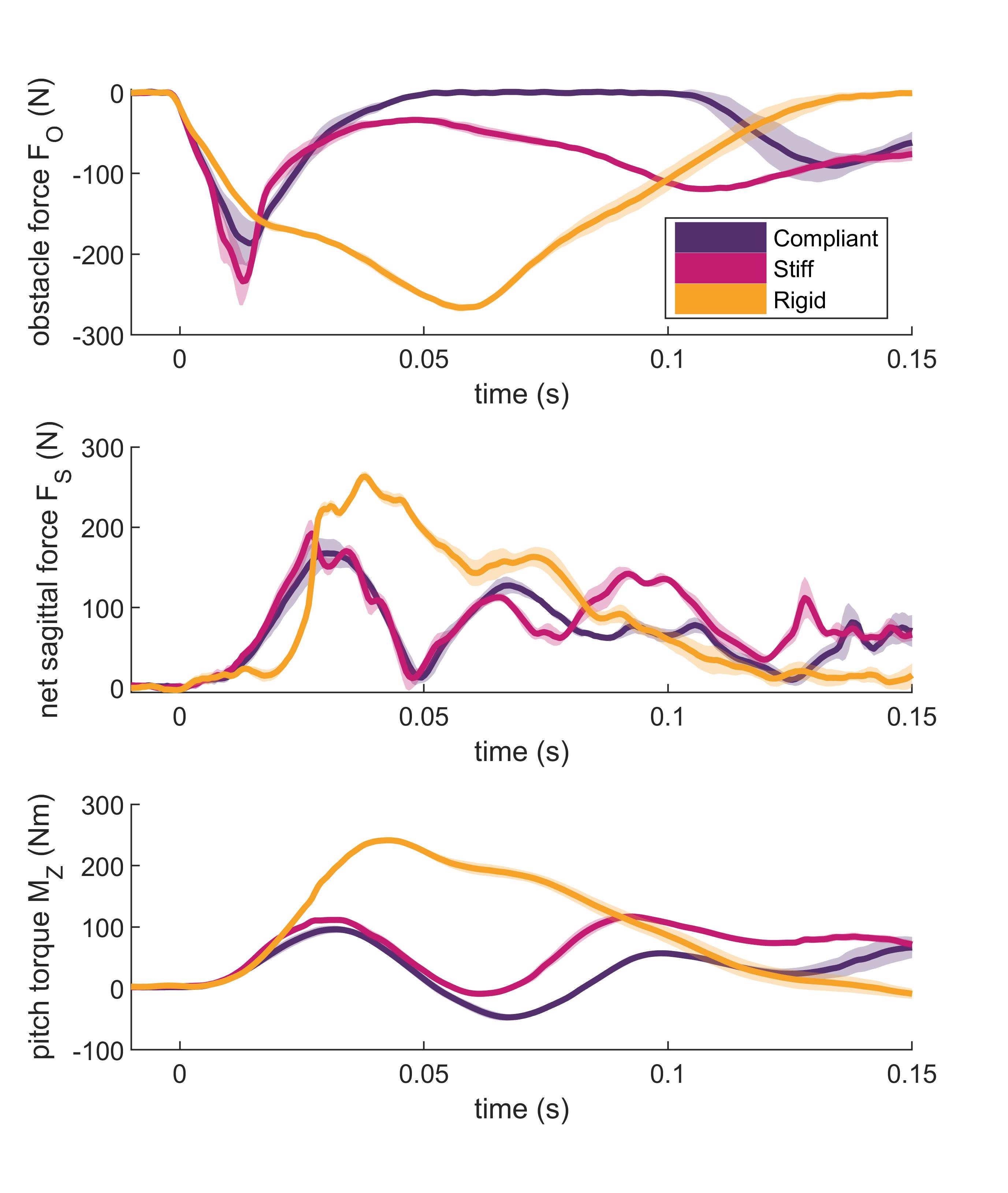}
    \caption[Collision force at the obstacle and interaction force and torque at the hip]{\textbf{Collision force at the obstacle and interaction force and torque at the hip} The average force and torque trajectories were repeatable as suggested by the standard deviation bands. A higher  maximum torque at the pelvic level can be observed with the RIGID knee (bottom). The maximum collision force at the obstacle varied with knee joint configuration and the peak was achieved later with the RIGID knee (top). Forces in the sagittal plane were slightly higher when the knee was RIGID (middle).}
    \label{fig:force_plots}
\end{figure}

\begin{figure}[htb]
    \centering
    \includegraphics[width=\textwidth/2]{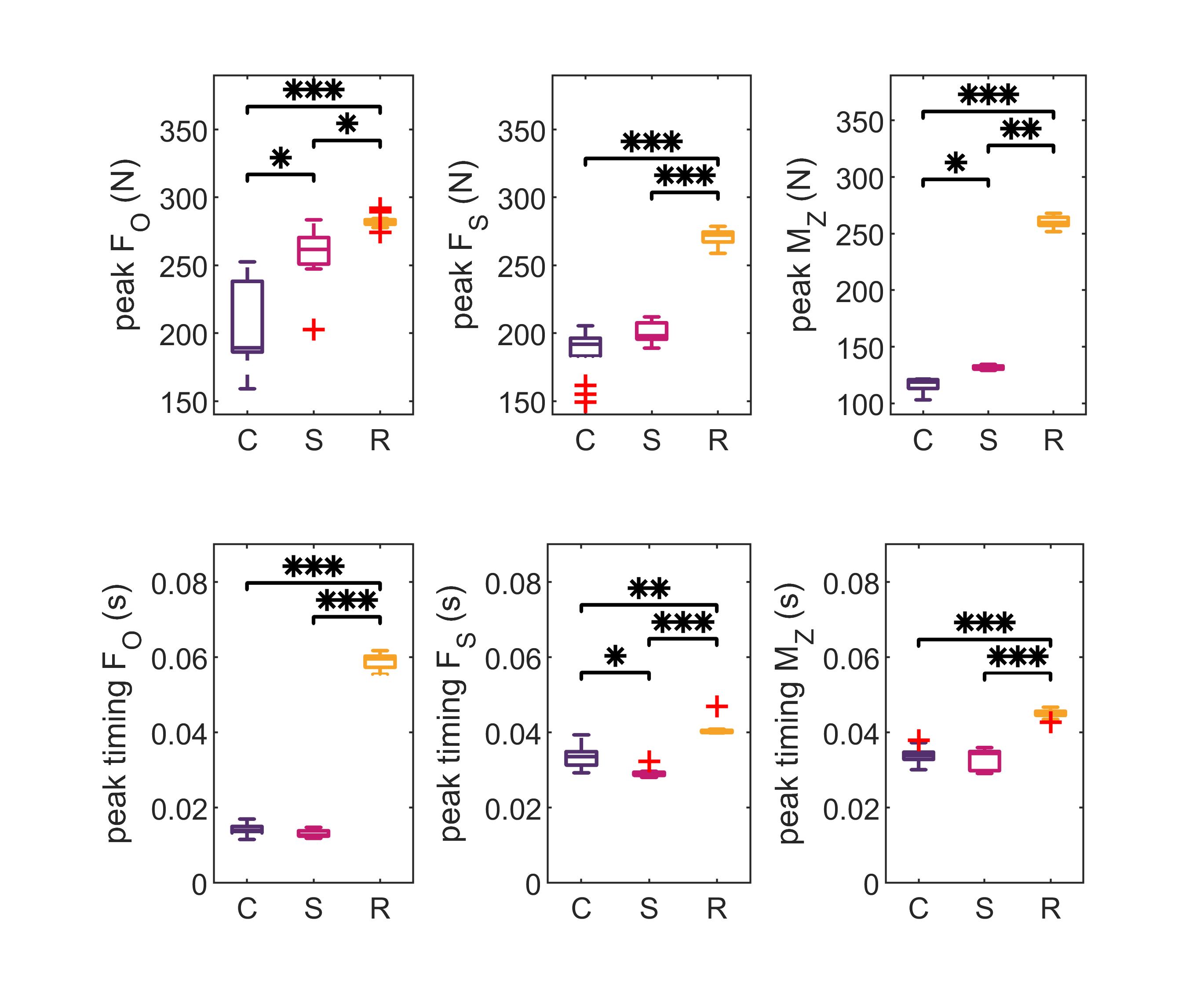}
    \caption[Peak values and peak timing after collision]
    {\textbf{Peak forces and torques during the collision} Peak amplitude (top) and timing (bottom) of force at the obstacle (left), sagittal force (middle) and pitch torque (right) with COMPLIANT (C), STIFF (S) and RIGID (R) joint configurations. Median, 25th and 75th percentiles, maximum and minimum values, and outliers are indicated by red line, box edges, whiskers and red crosses, respectively.  Statistically significant differences are indicated with a bracket while one, two and three stars indicate p-values lower than 0.05, 0.01 and 0.001, respectively}
    \label{fig:boxplot_peaks}
\end{figure}

The statistical analysis of the outcome parameters are presented in Tab.~\ref{tab:stat_results}. The different levels of statistical significance are also indicated in the boxplots with brackets. The average peak force at the obstacle changed from around 200 (COMPLIANT), to 260 (STIFF) and finally to 280~N (RIGID)~(Fig.~\ref{fig:force_plots} and Fig.~\ref{fig:boxplot_peaks}). The average net sagittal peak force was higher for the RIGID knee joint (271.2$\pm$5.5~N) compared to the other two configurations with 200.9$\pm$7.6~N and 185.9$\pm$17.1~N. Pitch torque ($M_Z$) values were approximately twice as high for the RIGID knee joint at 260.2$\pm$5.0~Nm compared to 132.0$\pm$1.8~Nm and 116.2$\pm$5.8~Nm for the STIFF and COMPLIANT knee, respectively. The timing of the peak force and torque at the pelvic level was similar (around 0.04~s) for all three knee joint configurations. The peak force at the obstacle, however, occurred at around 0.02~s for the COMPLIANT and STIFF knee joints, but only after 0.06~s with the RIGID knee~(Fig.~\ref{fig:boxplot_peaks}).

\begin{figure}[htb]
    \centering
    \includegraphics[width=\textwidth/2]{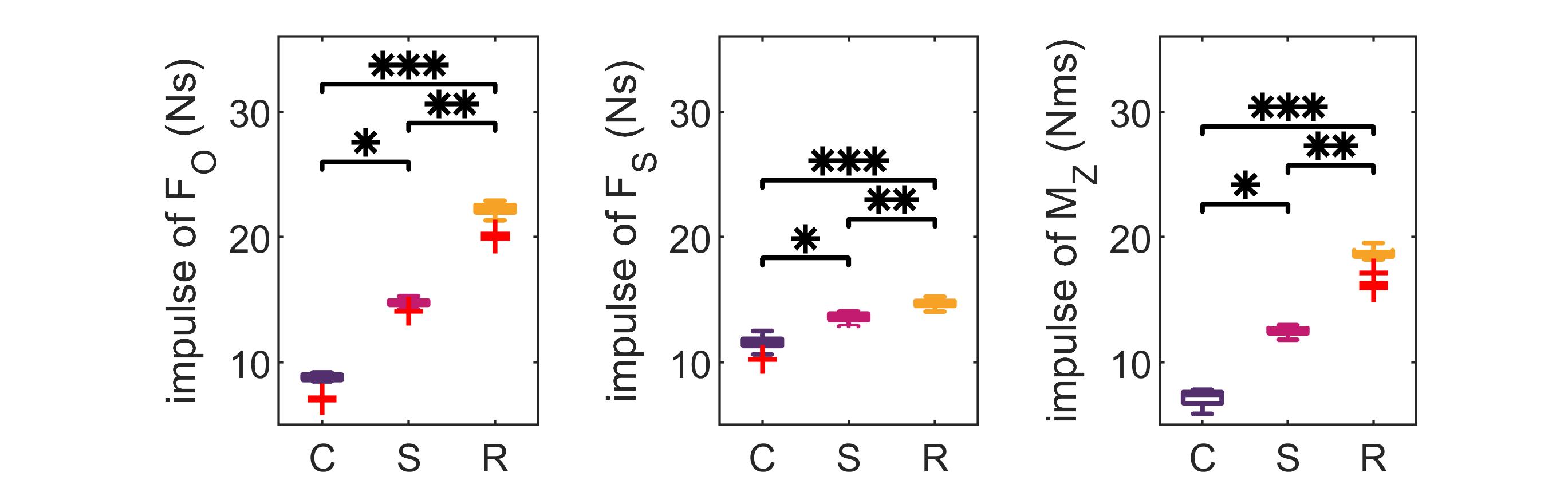}
    \caption[Collision impulse]{\textbf{Collision impulse} For the RIGID knee joint, the impulse was clearly highest for all parameters. The COMPLIANT and STIFF VSA configurations showed a bigger difference in impulse of the obstacle force $F_O$ and the torque at the hip $M_Z$. In contrast, the difference in impulse was less prominent for the sagittal plane force $F_S$ when comparing the two VSA configurations. Statistically significant differences are indicated with a bracket while one, two and three stars indicate p-values lower than 0.05, 0.01 and 0.001, respectively.}
    \label{fig:boxplot_impulse}
\end{figure}

The impulse generally increased with joint stiffness (Fig.~\ref{fig:boxplot_impulse}). For the pitch torque, the COMPLIANT knee resulted in an impulse three times smaller than the RIGID knee joint, and twice as small as the STIFF knee joint. The differences between the configurations in the obstacle force impulse were similar to the pitch torques. In contrast, the difference between the configurations in the impulse of the sagittal plane force was not as large. The incremental increase in the sagittal plane force between the three conditions was around 2~Ns, which corresponds to about 18\% of the impulse observed with the COMPLIANT knee joint.

%%%%%%%%%%%%%%%%%%%%%%%%%%%%%%%%%%%%%%%%%%%%%%%%%%%%%%%%%%%%%%%%%%%%%%%%%%%%%%%%
% ------ DISCUSSION ------------------------------------------------------------
%%%%%%%%%%%%%%%%%%%%%%%%%%%%%%%%%%%%%%%%%%%%%%%%%%%%%%%%%%%%%%%%%%%%%%%%%%%%%%%%

\section{Discussion}
This work aimed to examine the influence of knee joint stiffness of an exoskeleton leg on interaction forces and torques resulting from a collision during swing phase. Test bench measurements were conducted to compare the VSA in the VariLeg exoskeleton to a rigid actuation paradigm~\cite{Schrade2018}. As expected, the results showed that peak torques, and forces were reduced at the pelvic level when the knee joint was compliant. Additionally, the impulse for all observed forces and torques was reduced when the knee joint was compliant.

The noticeable decrease in peak force and torque, and impulse transmitted to the user observed with the COMPLIANT joint would increase the likelihood of the user being able to counteract these forces and torques with the walking aids. This could prevent falls and increase safety. Furthermore, the risk to injure the shoulder joints by sudden force and torque peaks during balancing would also be decreased. Lastly, decreased forces and torques acting on the exoskeleton would also decrease the risk of damaging the hardware.

Peak force at the obstacle varied with joint stiffness although theoretical models suggest that the stiffness of the joint should not influence the initial forces of an impact~\cite{Haddadin2009a}. This could be due to different collision behavior between RIGID joint and VSA joint configurations, as suggested by the different kinematic patterns. From the end-point trajectory, we observed that the COMPLIANT and STIFF knee joints allowed the leg to start lifting the end-point over the obstacle with continued knee and hip flexion after the collision. This is similar to what has been observed in experiments where prosthesis users were tripped during late swing phase~\cite{Shirota2015}. The RIGID knee joint, on the other hand, bounced back from the obstacle as knee flexion was prevented by the rigidity.

The fact that the leg started to climb over the obstacle despite not being programmed to do so may be beneficial for implementing collision mitigation strategies. A potential strategy to mitigate the effects of stumbling could be to try to lift the leg over the obstacle as humans do~\cite{Schillings2000}. When the leg just bounces off the obstacle and moves backwards, the inertia of the system delays adequate reaction. In that case, all the kinetic energy would have to be redirected upwards to step over the obstacle or towards the ground, which would increase the risk of falling. If the mechanics of the leg already start to redirect the kinetic energy into the correct direction (upwards or downwards), the stumbling mitigation can be achieved more quickly than if the leg were moving backwards.

In contrast to our expectations based on~\cite{Haddadin2009}, the timing of the peaks at the pelvic level was not notably delayed at lower levels of knee joint stiffness. At the obstacle however, the RIGID configuration exhibited a later occurrence of the peak collision force compared to the COMPLIANT and STIFF knee joint configurations. We hypothesize that this is due to the differing collision behavior. With the RIGID configuration, the tennis ball is deformed in an elastic collision and kinetic energy is redirected from a forward movement to a backwards movement. This causes higher penetration depth (more tennis ball deformation) and high forces are transmitted to the obstacle for a longer time. This hypothesis is supported by the impulse that was three times  higher for the pitch torque~($M_z$) with the RIGID knee joint configuration compared to the COMPLIANT knee joint.

It is interesting to note that the difference in peak pitch torque was only about 16~Nm comparing the COMPLIANT and STIFF configurations, which is about 10\% of the torque observed with the COMPLIANT knee joint. However, the difference in impulse was much larger as the STIFF knee joint resulted in double the impulse of the COMPLIANT knee joint. Accordingly, lowering the stiffness of the VSA during swing seems to be mostly relevant to reduce the impulse rather than peak forces and torques caused by collisions. Thus, an SEA offering a fixed mechanical compliance and rendering a broader range via control, similar to the ones used in other legged robots~\cite{Hutter2016}, may be sufficient in terms of reducing peak forces and torques. This would allow to reduce the complexity of the actuation system as the motor used to pretension the spring could be omitted, rendering the design lighter. Having some mechanical compliance in the system is most likely necessary, however, as even with fast sensing the collision has to be detected first and then acted upon. Actuation system inertia, which is typically high due to the large reduction ratios used in powered lower limb exoskeletons, would not allow adequate reaction to reduce the peaks observed here that occur during the first 150~ms of a collision. 

In the real application scenario, feed-back control would add disturbance rejection and error minimization. For our experimental protocol however, feed-back control would have led to results that are partially caused by the mechanical structure and partially by the implemented control. Hence, we only used the first layer of control consisting of the feed-forward part. This allowed slight differences in swing phase execution, which could be observed between the configurations. The slightly higher collision point observed with the RIGID knee joint is assumed to not have influenced the results as only the velocity is relevant for the kinetic energy at collision. Velocity differences would be more problematic than differing collision height as the velocity squared directly relates to the impulse over the kinetic energy in elastic collisions. Accordingly, comparing the total velocity of the end point at collision, the difference between knee joint configurations seemed negligible (0.05~m/s). A different approach would have been to move the obstacle to induce collision at exactly the point of maximum forward velocity of the leg. However, the risk of introducing more variability through moving the obstacle between configurations led us to rather accept this inaccuracy. Only considering the brief window after collision onset to investigate collisions also ensured that manually stopping the exoskeleton by cutting the motors off the power supply should not have influenced the results. 

The experiments were performed on a test bench only, with no user attached to the exoskeleton leg. A human user may significantly change the system behavior as passive mass, spring and damper elements are added in parallel to the exoskeleton leg (as a first approximation for non-spastic motor complete spinal cord injured legs). In addition, the torso and arms would react to the collision and interact with the environment. The forces that the user would have had to counteract to not fall over the obstacle were significantly lower in the case of the COMPLIANT knee joint as the interaction torque at the hip was only half as big compared to the RIGID knee joint. Further, we argue that the lower impulse with the COMPLIANT knee joint indicated that the overall effort to counteract collision forces would be lower for a user since lower impulse means that less force or torque had to be exerted during the collision time window.

It would be interesting to perform similar experiments with users inside the exoskeleton. However, this would likely increase the variability of the data as the human user might react differently every time. Although non-trivial, this has already been investigated in prosthesis users~\cite{Shirota2014}. Besides the technical difficulties, such an experiment would introduce the risk of falling, straining muscles or similar that need to be taken into consideration. 

A notable difference between amputees and motor complete SCI users of exoskeletons would be that SCI users have no sound limb to rely on for recovery~\cite{Shirota2015}. Thus, the exoskeleton would have to provide all the recovery support. An advantage of using series elasticity is that the elastic elements can be used as torque sensors. Deviations of the joint from its equilibrium position during swing phase can be observed to estimate the torque acting on the joint. Stumbling mitigation strategies could be initiated if the deviation from the equilibrium position is larger than a predefined threshold, thus improving fall protection capabilities of exoskeletons. A collision detection algorithm similar to~\cite{DeLuca2006} could be implemented with a series elasticity detecting onset of collisions without additional torque sensors. After detection of the collision, the controller could react to it by stepping over the obstacle or trying to prematurely end the swing phase and placing the foot on the floor to step over the obstacle with the other leg, similar to what has been developed for prostheses~\cite{Lawson2010}. Testing such algorithms on the same test bench setup would yield valuable insight towards better fall prevention.

%%%%%%%%%%%%%%%%%%%%%%%%%%%%%%%%%%%%%%%%%%%%%%%%%%%%%%%%%%%%%%%%%%%%%%%%%%%%%%%%
% ------ CONCLUSION ------------------------------------------------------------
%%%%%%%%%%%%%%%%%%%%%%%%%%%%%%%%%%%%%%%%%%%%%%%%%%%%%%%%%%%%%%%%%%%%%%%%%%%%%%%%

\section{Conclusion}
\label{sec:conclusion}
Adding compliance to the knee of a powered lower-limb exoskeleton leg reduced peak forces and torques at the pelvic level in a test bench collision experiment. Even more, it reduced the impact impulse during the collision, which could reduce the effort a user has to exert to maintain balance. Further lowering stiffness with the VSA during swing phase noticeably reduced impulse, and, to a lesser extent, peak torques and forces at the pelvic level. This suggests that compliance in the range suggested here lead to safer collisions and variations in the VSA stiffness range investigated are mostly beneficial to reduce impulse. These findings should inform future designs of wearable lower limb exoskeletons and can help to design joints that are mechanically inherently robust to collisions. This may bring the field one step closer to robots that move as nimbly and swiftly as animals and humans.

%%%%%%%%%%%%%%%%%%%%%%%%%%%%%%%%%%%%%%%%%%%%%%%%%%%%%%%%%%%%%%%%%%%%%%%%%%%%%%%%
% ------ APPENDIX --------------------------------------------------------------
%%%%%%%%%%%%%%%%%%%%%%%%%%%%%%%%%%%%%%%%%%%%%%%%%%%%%%%%%%%%%%%%%%%%%%%%%%%%%%%%

%\addtolength{\textheight}{-4.1cm} 
\section{Appendix: Trajectory Generation and Following}
\label{Appendix:Control}
\begin{figure}[ht]
      \centering
     \includegraphics[trim={0.6cm 0cm 0.55cm 0cm},clip,width=0.9\columnwidth]{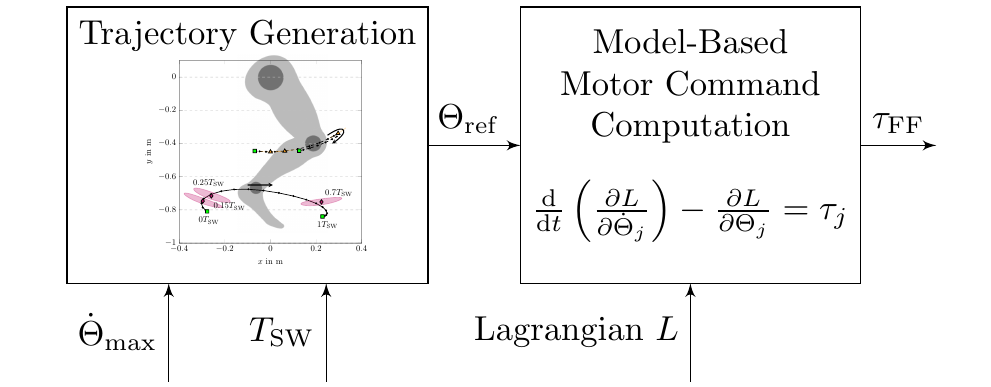}
      \caption{Two step computation of feedforward motor commands.}
      \label{fig:calc_database}
\end{figure}

The approach for computing nominal motor commands (feed-forward control) consist of two steps (Figure~\ref{fig:calc_database}):
First, trajectories of the swing phase are generated considering the speed limit of the motors $\dot \Theta_{\rm max}$ and the desired swing phase duration $T_{\rm SW}$.
%This step is discussed in Section~\ref{sec:traj_gen}.
Second, the motor commands which are necessary to follow this swing phase trajectory are computed.
Limitations of available motor torques are checked in this step.
%This step utilizes Lagrangian mechanics and is discussed in Section~\ref{sec:motorbased}.
The objective is to find motor commands for the three motors such that the reference trajectory for the hip angle, the knee angle, and the pretension for the spring of the VSA, i.e. the winch motor angle, is followed.

\subsection{Trajectory Generation}
\label{sec:traj_gen}
%We first outline our optimization-based approach to compute the reference trajectories, i.e. the reference angles and angular velocities of the hip and the knee joints.
The reference angles and angular velocities of both the hip and the knee joints were obtained using constrained optimization.
The optimization problem was formulated to consider the speed limit of the motors, to ensure ground clearance, and to minimize the maximum curvature of the angular velocity trajectories of both the hip angle and the knee angle. 
The resulting optimization problem is convex and is given by
\begin{subequations}
\label{eq:opt}
\begin{alignat}{2}
&\underset{\phi_i,\dot \phi_i,\beta_i,\dot \beta_i}{\text{minimize}}        &\quad & \max_i\{{\rm curv}(\dot \phi_i),{\rm curv}(\dot \beta_i)\}\label{eq:objective}\\
& \begin{matrix} \text{subject to} \\ \hspace{0.1cm} \end{matrix} 
                    &      
                    &
\begin{matrix}
\phi_{i+1}-\phi_{i}=T_s \dot \phi_{i}  
\\
\beta_{i+1}-\beta_{i}=T_s \dot \beta_{i}
\end{matrix}
\qquad  \forall i=0,...  T_{\rm st}
 \label{eq:constraint1}
 \\
 &
                    &      &
                    \begin{matrix}
                     |\dot \phi_{i}|\leq \theta_{\rm max} 
                     \\
                      |\dot \beta_{i}|\leq \theta_{\rm max} 
					\end{matrix}                     
                     \qquad  \forall i=0,...  T_{\rm st} \label{eq:constraint2}
 \\
&                  &      & \left\| 
\begin{bmatrix}
\phi_{T_j} %\\ \chi_{T_j}
%\end{bmatrix}
- 
%\begin{bmatrix}
\phi_{T_j,\rm ref} \\ 
\beta_{T_j}- \beta_{T_j, \rm ref}
\end{bmatrix}
\right\|_2^2 
\leq 5^\circ \quad j=1,2,3 \label{eq:constraint3}\\
&                  &      & 
\begin{bmatrix}
\phi_{T_j} \\ 
\beta_{T_j} 
\end{bmatrix}
=
\begin{bmatrix}
\phi_{T_j,\rm ref} 
\\
\beta_{T_j,\rm ref}
\end{bmatrix} \quad  j=0,4\label{eq:constraint4}\\
& & & 
\begin{bmatrix}
\dot \phi_{T_0} \\ 
\dot \beta_{T_0} 
\end{bmatrix}
=\begin{bmatrix}
0 \\ 
0
\end{bmatrix},\
\begin{bmatrix}
| \dot \phi_{T_4} | \\ 
| \dot \beta_{T_4} |
\end{bmatrix}
\leq 
\begin{bmatrix}
5^\circ/s \\ 
5^\circ/s 
\end{bmatrix}\label{eq:constraint5}
\end{alignat}
\end{subequations}
where $\phi_i$ is the hip angle, $\dot \phi_i$ is the angular velocity of the hip, $\beta_i$ is the knee angle, and $\dot \beta_i$ is the angular velocity of the knee at time $i$. 
 ${\rm curv}(\dot \theta_i)$ is the curvature of the angular velocity in angle space defined as fourth order central finite difference:
\begin{align*}
{\rm curv}(\dot \theta_i)
:=\frac{- \dot \theta_{i-2}+16\dot \theta_{i-1} - 30 \dot \theta_{i}+16\dot \theta_{i+1}- \dot \theta_{i+2}}{12}.
\end{align*}
This objective \eqref{eq:objective} is motivated by minimizing abrupt changes of the torque inputs that are necessary to follow the reference trajectory.
%The objective minimizes the maximum curvature over the execution.
%The angular acceleration relates to the torques.
%Hence, the third order derivative of the angle (== curvature of angular velocity) relates to torque changes.
Angles and their velocities are related through \eqref{eq:constraint1} with sampling time $T_s$;
constraint~\eqref{eq:constraint2} ensures that the speed limits of the motors are respected;
\eqref{eq:constraint3} constrains the angles at the specific time $T_j$ to remain within a circle of $5^\circ$ around pre-specified reference angles $\theta_{T_j,\rm ref}$ (in order to ensure ground clearance), specified in Table~\ref{tb:refs};
%The latter ensures ground clearance in the execution of the swing trajectory.
constraint~\eqref{eq:constraint4} fixes the initial and end positions of the swing phase trajectory; and \eqref{eq:constraint5} constrains the initial and end velocities.
Figure~\ref{fig:traj_gen} visualizes constraint~\eqref{eq:constraint3} with diamond markers and gray-shaded areas and \eqref{eq:constraint4} with square markers in task space.
The black curve visualizes the reference trajectory as minimizer of \eqref{eq:opt}.
\begin{table}[h]
\setlength\tabcolsep{8.2pt}
\caption{
Waypoints of swing trajectory
}
\begin{center}
\def\arraystretch{1.025}
\label{tb:refs}
\begin{tabular}{llllll llll}
\hline
Time in $T_{\rm SW}$ & 
$0$& 
$0.15$& 
$0.25$& 
$0.7$ &
$1$
 \\
\hline
\hline
Hip Angle $\phi_{T_j,\rm ref}$ &  $-9^\circ$ & $0^\circ$ & $8^\circ$ & $41^\circ$  & $16^\circ$ \\
Knee Angle $\beta_{T_j,\rm ref}$ &  $21^\circ$ & $45^\circ$ & $58^\circ$ & $51^\circ$ & $2^\circ$ \\
\hline
\end{tabular}
\end{center}
\end{table}

\subsection{Model-Based Motor Command Computation}
\label{sec:motorbased}

We utilize Lagrangian mechanics in order to derive the system dynamics with the state vector
$$
\Theta(t) = \begin{bmatrix}
\phi(t)
&
\chi(t)
&
\psi(t)
&
\alpha(t)
\end{bmatrix}^T,
$$ 
with the hip angle $\phi(t)$, lever angle $\chi(t)$, winch angle $\psi(t)$, and the deflection angle of the VSA $\alpha(t)$,
where $\beta(t)=\chi(t)+\alpha(t)$ is the knee angle.
The Lagrangian
\begin{align}
\label{eq:lagrangian}
L(t)=T(t)-V(t)
\end{align}
is defined in terms of the kinetic energy $T(t)$ and the potential energy $V(t)$:
\begin{align*}
 T(t) =\ & 
 \left( 
I_E ({\dot \phi}^2(t) + {\dot \chi}^2(t)+ {\dot \psi}^2(t)) \right.
   +m_u v_u^2(t) +m_l v_l^2 (t)
\\
& \hspace{0.6cm} \left. + I_u {\dot \phi}^2(t) +I_l (\dot \phi(t) -\dot \chi(t) -\dot \alpha(t)) ^2
 \right)/2
\\
 V(t) =\ & m_u gy_u(t)+m_lgy_l(t) +\frac{1}{2}k_\theta(t)\alpha^2(t)+\frac{1}{2}k \Delta x^2(t)
\end{align*}
with spring pretension $\Delta x(t) = r_w \psi(t)$
and
\begin{align*}
y_u(t)=\ &-r_{th} \cos \phi(t)
\\
 y_l(t) =\ & -l_{th} \cos \phi(t) - r_{sh} \cos (\phi(t) - \chi(t) -\alpha(t))
\\
 v_{th}^2(t) =\ & r_{th}^2 {\dot \phi}^2(t)
\\
 v_{sh}^2(t) =\ & l_{th}^2 {\dot \phi}^2(t)+ r_{sh}^2 {(\dot \phi(t)-\dot \chi(t) -\dot \alpha(t))}^2
 \\ &
+ 2l_{th} r_{sh} \dot \phi(t) (\dot \phi(t) -\dot \chi(t)-\dot \alpha(t))\cos \chi(t).
\end{align*}
The variable knee joint stiffness $k_\theta(t)$ is given by 
\begin{align*}
 k_\theta(t) =\ & BCk\cos\alpha(t)
 \left(
 \frac{\Delta x - |B-C|}{
 \sqrt{B^2+C^2-2BC\cos\alpha(t)}
 }+1\right)
 \\
 &
 -
 B^2C^2k\sin^2\alpha(t)
 \frac{\Delta x - |B-C|}{ (B^2+C^2-BC\cos\alpha(t))^{3/2}}
\end{align*}
in accordance with the model introduced by Van~Ham~et~al.~\cite{VanHam2007} and the model we experimentally validated in another piece of work~\cite{Schrade2018}. 
The model parameters are defined in Table~\ref{tb:params}.

\begin{table}[h]
\setlength\tabcolsep{6.2pt}
\caption{
Model Parameters
}
\begin{center}
\def\arraystretch{1.025}
\label{tb:params}
\begin{tabular}{lllll llll}
\hline
Symbol & Parameter & value 
 \\
\hline
\hline
 $r_{th}$ & center of gravity thigh & $0.243$~m \\
 $l_{th}$ & distance hip to knee & $0.485$~m \\
 $r_{sh}$ & center of gravity shank & $0.225$~m \\
 $r_w$ & radius of winch & $0.015$~m \\
 $m_{th}$ & mass thigh & $9.7$~kg \\
 $m_{sh}$ & mass shank & $4.2$~kg \\
 $g$ & gravitational acceleration & $9.81$~m/s${}^2$ \\
 $I_u$ & rotational inertia thigh & $0.222$~kgm${}^2$ \\
 $I_l$ & rotational inertia shank & $0.0712$~kgm${}^2$ \\
 $I_E$ & rotational inertia motors & $8.556$~kgm${}^2$ \\
 $k$ & stiffness of linear spring & $109$~kN/m \\
 $B$ & MACCEPA lever arm & 0.075~m \\
 $C$ & distance from knee$_{CoR}$ to winch pulley & 0.3~m \\
\hline
\end{tabular}
\end{center}
\end{table}

The system dynamics using the Lagrangian in \eqref{eq:lagrangian} yields
\begin{align}
\label{eq:dynamics}
    \frac{\rm d}{\rm dt}
    \frac{\partial L(t)}{\partial \dot \Theta(t)}
    -
    \frac{\partial L(t)}{\partial \Theta(t)}
    =
    \begin{bmatrix}
    \tau_h(t) - \tau_v(\dot \phi(t)
    \\
    \tau_k(t)- \tau_v(\dot \chi(t))
    \\
    \tau_w(t) - \tau_v(\dot \psi(t)) 
    \\
    0
    \end{bmatrix}
\end{align}

with the motor torques of the hip $\tau_h(t)$, knee $\tau_k(t)$, and winch $\tau_w(t)$, the viscous friction torque $\tau_v(\cdot)$.

%%%%%%%%%%%%%%%%%%%%%%%%%%%%%%%%%%%%%%%%%%%%%%%%%%%%%%%%%%%%%%%%%%%%%%%%%%%%%%%%
% ------ ACKNOWLEDGMENT --------------------------------------------------------
%%%%%%%%%%%%%%%%%%%%%%%%%%%%%%%%%%%%%%%%%%%%%%%%%%%%%%%%%%%%%%%%%%%%%%%%%%%%%%%%

\section*{Acknowledgment}
This work was supported by the Swiss National Science Foundation through the National Centre of Competence in Research on Robotics, the ETH Research Grant ETH-22 13-2, the ETH Zurich Foundation in collaboration with Hocoma~AG, Swiss National Science Foundation, grant no. PP00P2 157601 /1 and the EMPA~(division 304, mechanical systems engineering).

We gratefully acknowledge the  support of Bernhard Weisse of EMPA Switzerland and that of Kistler AG Switzerland. We also thank Michael Baumann, Samuel Bianchi and Frank Grossenbacher for preparing the test bench.

%%%%%%%%%%%%%%%%%%%%%%%%%%%%%%%%%%%%%%%%%%%%%%%%%%%%%%%%%%%%%%%%%%%%%%%%%%%%%%%%
% ------ BIBLIOGRAPHY --------------------------------------------------------------
%%%%%%%%%%%%%%%%%%%%%%%%%%%%%%%%%%%%%%%%%%%%%%%%%%%%%%%%%%%%%%%%%%%%%%%%%%%%%%%%

%\bibliographystyle{no-dash_Ieeetran}
\bibliographystyle{IEEEtran}
\bibliography{IEEEabrv,main}

\end{document}